\documentclass[conference]{IEEEtran}
\IEEEoverridecommandlockouts
\usepackage[noadjust]{cite}
\usepackage{amsmath,amssymb,amsfonts, bm}
\usepackage{algorithmic}
\usepackage{graphicx}
\usepackage{textcomp}
\usepackage{xcolor}
\usepackage{comment}
\usepackage{mathptmx}
\usepackage{amsthm}
\usepackage{nicefrac}
\usepackage{dsfont}
\usepackage{authblk}
\usepackage{etoolbox}    
\usepackage{dirtytalk}
\usepackage{changepage}
\usepackage{enumitem}
\usepackage{cuted}
\usepackage{refcount}
\usepackage{mathtools}
\usepackage{optidef}
\usepackage{url}
\usepackage{stfloats}
\usepackage[cal=cm]{mathalfa}
\newtheorem{thm}{Theorem}
\newtheorem{lem}{Lemma}
\newtheorem{prop}{Proposition}

\theoremstyle{definition}
\newtheorem{defn}{Definition}

\def\BibTeX{{\rm B\kern-.05em{\sc i\kern-.025em b}\kern-.08em
    T\kern-.1667em\lower.7ex\hbox{E}\kern-.125emX}}

\usepackage{bbm}

\newcommand{\customtag}{\refstepcounter{equation}\tag{\the\numexpr\getrefnumber{eq: lmmse estimate}+1}}

\DeclareMathOperator*{\argmin}{\arg\!\min}
\newenvironment{sketchproof}{%
  \proof}{\endproof}

\definecolor{NYUviolet}{HTML}{57068c} 	
\definecolor{NYUlight}{HTML}{8900e1} 	
\definecolor{NYUdark}{HTML}{330662} 	
\definecolor{NYUnight}{HTML}{220337} 	

\newtoggle{shortversion}
\togglefalse{shortversion}

\begin{document}

\title{
 Pilot-Attacks Can Enable Positive-Rate Covert Communications of Wireless Hardware Trojans
\thanks{This work is supported by National Science Foundation grants 1815821 and 2148293.}} 

\author[1]{Serhat Bakirtas}
\author[2]{Matthieu R. Bloch}
\author[1]{Elza Erkip}

\affil[1]{Tandon School of Engineering, New York University, Brooklyn, NY \authorcr
  \{serhat.bakirtas, elza\}@nyu.edu}
\affil[2]{School of Electrical and Computer Engineering, Georgia Institute of Technology, Atlanta, GA \authorcr
  matthieu.bloch@ece.gatech.edu}

\maketitle

\begin{abstract}
Hardware Trojans can inflict harm on wireless networks by exploiting the link margins inherent in communication systems. We investigate a setting in which, alongside a legitimate communication link, a hardware Trojan embedded in the legitimate transmitter attempts to establish communication with its intended rogue receiver. To illustrate the susceptibility of wireless networks against pilot attacks, we examine a two-phased scenario. In the channel estimation phase, the Trojan carries out a covert pilot scaling attack to corrupt the channel estimation of the legitimate receiver. Subsequently, in the communication phase, the Trojan exploits the ensuing imperfect channel estimation to covertly communicate with its receiver. By analyzing the corresponding hypothesis tests conducted by the legitimate receiver in both phases, we establish that the pilot scaling attack allows the Trojan to operate in the so-called \say{linear regime} i.e., covertly and reliably transmitting at a positive rate to the rogue receiver. Our results highlight the vulnerability of the channel estimation process in wireless communication systems against hardware Trojans.
\end{abstract}
\begin{IEEEkeywords}
hardware Trojans, wireless communications, covert communications, pilot corruption attack
\end{IEEEkeywords}

\section{Introduction}
\label{sec: introduction}

Assuring confidentiality, integrity, and authenticity of transmissions in communication networks has always been of prime importance. Recently, however, the concept of achieving a low probability of detection, or \emph{covertness}, has garnered increased attention~\cite{Bash2013}. This renewed interest is partly driven by the understanding that the mere knowledge of a party’s communication can be as significant as the content of the communication itself. This interest also stems from concerns about potential side channels that could surreptitiously exfiltrate sensitive information~\cite{Sangodoyin2021Side}. The present work is particularly motivated by the latter concern, focusing on the opportunities that hardware Trojans have to exist “in the margins.” These margins are inherent in communication protocols, designed to accommodate minor imperfections and variability~\cite{subramani2019demonstrating, subramani2020amplitude}. Given the ubiquity of pilot symbols in contemporary wireless protocols, this study aims to explore the feasibility and impact of attacks where hardware Trojans manipulate the pilot symbols in an undetected way. The ultimate goal is to understand how such manipulation could undermine the detection capabilities of monitoring entities in subsequent transmissions.


Theoretical explorations of covert capacity have unveiled two distinct regimes of covert communications. The first is the \emph{square-root law} regime~\cite{Bash2013,bloch2016covert,Wang2016b}, in which the number of covert bits must scale with the square root of the blocklength. The second is the \emph{linear} regime, in which the number of bits can scale linearly with the block length~\cite{lee2018covert}. Operating within the linear regime typically necessitates the exploitation of uncertainty in channel state knowledge~\cite{Che2014a,Sobers2017,Lee2015,ZivariFard2020}. Specifically, the introduction of artificial noise is a potent signaling technique used to engineer this uncertainty~\cite{Tekin2008General}.


In the present work, we examine a scenario where a hardware Trojan manipulates pilot symbols with the intent to diminish the channel estimation accuracy of legitimate parties. This manipulation subsequently curtails their capacity to detect communication initiated by the hardware Trojan. A significant contribution of our research is the demonstration of how pilot symbol manipulation by a hardware Trojan can effectively bypass the square root law, thereby facilitating operation within the linear regime. This finding underscores the potential risks posed by hardware Trojans in modern communication systems.


The organization of the rest of this paper is as follows. In Section~\ref{sec: problem formulation} we formally introduce the system model. In Section~\ref{sec: main result}, we present our main results and their proof. Finally, in Section~\ref{sec: conclusion}, we offer concluding remarks and discuss future directions. The full proofs are provided in the longer version of this paper~\cite{longerversion}.

\noindent{\em Notation:} We denote scalars with lowercase letters, vectors with lowercase bold letters, and matrices with uppercase bold letters. For vectors $\|\cdot\|_2$ denotes the Euclidian norm and for matrices $|\cdot|$ denotes the determinant. $\mathcal{CN}(\mu,\sigma^2)$ denotes circularly-symmetric complex Gaussian distribution with respective mean $\mu$ and variance $\sigma^2$. $\mathbb{D}$ denotes the Kullback-Leibler divergence~\cite[Chapter 2.3]{cover2006elements}. $\mathcal{O}$, $o$, $\Theta$, and $\omega$ follow the standard Bachmann–Landau notation~\cite[Chapter 3]{cormen2022introduction}. Unless stated otherwise, $\log$ denotes the natural logarithm. Finally, $L$ and $n$ denote the pilot sequence length and communication block length, respectively, and for a sequence $\{r_{i}\}_{i=1}^{n+L}$, we refer to $\{r_{i}\}_{i=1}^L$ and $\{r_{i}\}_{i=L+1}^{n+L}$ by $\bm{r}^{\textnormal{est}}$ and $\bm{r}^{\textnormal{comm}}$, indicating channel estimation and communication phases, respectively.

\section{Problem Formulation}
\label{sec: problem formulation}

\begin{figure}[t]
\centerline{\includegraphics[width=0.35\textwidth,trim={19cm 8cm 17cm 0},clip]{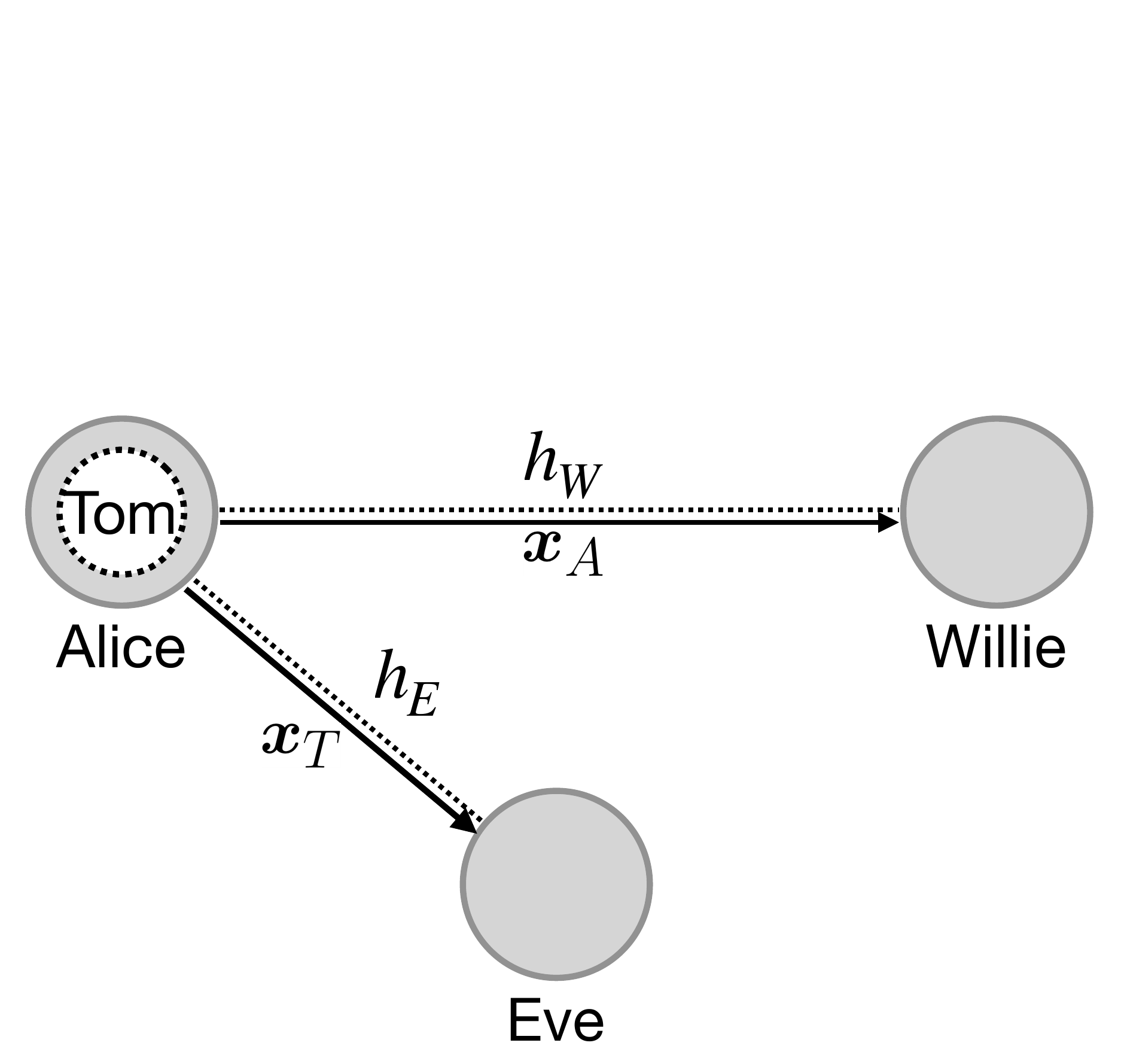}}
\caption{Legitimate transmitter, Alice, communicates with her intended (legitimate) receiver, Willie. Simultaneously, hardware Trojan, Tom, embedded in Alice, also communicates with his intended rogue receiver, Eve. Willie's objective is to decode Alice's signal $\bm{x}_A$ and detect the existence of any rogue signal $\bm{x}_T$.}
\label{fig: system model}
\end{figure}

As illustrated in Figure~\ref{fig: system model}, we consider a four-terminal scenario in which (1) a legitimate transmitter Alice, communicates with a legitimate receiver Willie; (2) a hardware Trojan Tom embedded in Alice, seeks to simultaneously communicate with a rogue receiver Eve, while evading detection by Willie, who also acts as a monitoring entity. 

We assume that Alice, Tom, Willie, and Eve each have a single antenna. We adopt a Rayleigh block fading channel model by which the received signal $\bm{y}_u$ at the user $u\in\{W,E\}$ is given by
\begin{align}
\begin{array}{ll}
    y_{u,i} = \alpha_u \: h_u \: x_i \: +\: z_{u,i}, & \forall i\in [n+L]
\end{array}
\label{eq: received signal general model}
\end{align}
where $\alpha_W$ and $h_W$ (resp. $\alpha_E$ and $h_E$) denote the propagation loss and the channel fading gain between Alice and Willie (resp. Tom and Eve). We assume that $h_u\sim \mathcal{CN}(0,\sigma_H^2)$ is independent of the transmitted sequence $\{x_i\}_{i=1}^{n+L}$ and noise $\{z_{u,i}\}_{i=1}^{n+L}$ and remains constant for at least $n+L$ symbol periods. We assume $\smash{z_{u,i}\overset{\textup{i.i.d.}}{\sim} \mathcal{CN}(0,\sigma_u^2)}$. Since Tom is embedded in Alice, $h_W$ and $\alpha_W$ (resp. $h_E$ and $\alpha_E$) also denote the channel gain and propagation loss between Tom and Willie (resp. Alice and Eve).

Motivated by practical wireless communication networks, our proposed system model comprises two distinct phases. In the first phase, called the \emph{channel estimation phase}, Alice sends a known pilot sequence for Willie to estimate the channel. In an effort to improve his chances at covertness in the subsequent phase, Tom attempts to covertly corrupt Willie's estimate by scaling the pilot sequence by $1+\epsilon$ for some small $\epsilon>0$ called the \emph{scaling parameter}. Simultaneously, Willie attempts to detect whether the pilot sequence is corrupted by scaling or not. In the second phase, called the \emph{communication phase}, Alice and Tom communicate with their respective receivers Willie and Eve, while Willie once again attempts to detect any rogue communication between Tom and Eve.

Willie's detection falls under the framework of simple binary hypothesis tests. We let the null hypothesis $H_0$ in the channel estimation phase correspond to the situation in which Alice's pilot sequence is not corrupted by Tom and the alternative hypothesis $H_1$ correspond to the situation in which Alice's pilot sequence is scaled by $(1+\epsilon)$ by Tom. Formally,
\begin{align}
\begin{array}{lll} H_0: & x_i = s_{A,i} &\forall i\in[L],\\
H_1: &x_i=(1+\epsilon) s_{A,i}  &\forall i\in[L], 
\end{array}\label{eq: transmitted pilot sequence}
\end{align}
where $\bm{s}_A$ is Alice's pilot sequence of length $L$, known to Tom, Eve and Willie. We assume $L=o(n)$ and $\|\bm{s}_A\|_2^2=\omega_L(1)$ to enable reliable channel estimation~\cite[Section III-A]{hassibi2003much}.

We let the null hypothesis $\widetilde{H}_0$ in the communication phase correspond to the situation in which the only transmission is between Alice and Willie, and the alternative hypothesis $\widetilde{H}_1$ correspond to the situation in which, in addition to the legitimate communication link between Alice and Willie, Tom also communicates with Eve. Formally,
\begin{align}
\begin{array}{lll}
     \widetilde{H}_0: &  x_{L+i}=x_{A,i} & \forall i\in[n],\\
     \widetilde{H}_1: &  x_{L+i}=x_{A,i}+x_{T,i} & \forall i\in[n],
\end{array}
\label{eq: transmitted signal}
\end{align}
where $\bm{x}_A$ and $\bm{x}_T$ denote Alice's and Tom's channel inputs, respectively.

We assume that $\bm{x}_A$ and $\bm{x}_T$ are mutually orthogonal zero-mean complex Gaussian sequences with i.i.d. components with a deterministic short-term~\cite{caire1999optimum} power constraint. Formally,
\begin{align}
    \frac{1}{n}\|\bm{x}_A\|_2^2 = \Lambda_A, \hspace{2em} &
    \frac{1}{n}\|\bm{x}_T\|_2^2 = \Lambda_T.
\end{align}
In addition, we assume that $\bm{x}_A$ and $\bm{x}_T$ are orthogonal to $\bm{z}^{\textnormal{comm}}$. This is justified by independence in the asymptotic regime as $n\to\infty$.

We further assume that both Alice and Tom know $h_W$ and $h_E$ perfectly. This, for example, could happen in a TDD system with channel reciprocity in which Eve is also a legitimate user and no Trojan is present at Willie or Eve to cause pilot corruption.

As argued in~\cite{subramani2019demonstrating,subramani2020amplitude}, the transmitters in typical wireless communication scenarios do not operate at the capacity because of design choices. Therefore, we assume that Alice adopts a link margin, transmitting at a rate strictly lower than the instantaneous channel capacity to Willie for given channel realization $h_W$ under a short-term power constraint. Formally, we assume that Alice transmits to Willie at a rate $R_A$ such that
\begin{align}
    R_A & < \log_2\left(1+\frac{\alpha_W^2 |h_W|^2 \Lambda_A}{\sigma_W^2}\right).\label{eq: alice's rate}
\end{align}
Throughout, we assume Alice and Tom use their knowledge of $h_W$ and $h_E$ to ensure no outage takes place in their respective communications. Hence, the instantaneous capacity from Alice to Willie given by the RHS of Eq.~\eqref{eq: alice's rate}, also known as the delay-limited capacity~\cite{hanly1998multiaccess}, is the appropriate bound for $R_A$.

The performance of any simple binary hypothesis test is captured by the trade-off between the false alarm probability $\mathbb{P}_F$ and the missed detection probability $\mathbb{P}_M$. Observe that Willie may always perform a \emph{blind test} ignoring his received signal and pick hypotheses based on an independent Bernoulli random variable, achieving $\mathbb{P}_F+\mathbb{P}_M=1$ in either phase. Hence, as is customary in the literature~\cite{Bash2013,bloch2016covert,ta2019covert,lee2015achieving}, Tom's covertness objective is to make Willie's detection strategy comparable to a blind test. 

For tractability, we assume that Willie performs distinct hypothesis tests in the different phases. Specifically, if Willie's test does not perform substantially better than a blind test in the channel estimation phase (Eq.~\eqref{eq: covertness criterion 1}), Willie fails to detect Tom and acts based on the null hypothesis $H_0$ in the communication phase (Eq.~\eqref{eq: willie's assumption}).
Thus, Tom's subsequent objective becomes preying on Willie's initial failure and communicating covertly to Eve (Eq.~\eqref{eq: covertness criterion 2}). While doing so, Tom's actions should not disrupt successful decoding in the legitimate Alice-Willie link (Eq.~\eqref{eq: covertness criterion 3}). Tom's communication commences only if Willie fails to detect the Trojan in the channel estimation phase. 

Our covertness criteria are formally defined in Definition~\ref{defn: covertness criterion} below. 

\begin{defn}{\textbf{(Covertness Criteria)}}\label{defn: covertness criterion}
Given a \emph{detection budget} $(\delta_1,\delta_2)$ and Alice's transmit power and rate  pair $(\Lambda_A,R_A)$, Tom remains \emph{covert} if
    \begin{align}
    &\lim\limits_{L\to\infty} \mathbb{P}^{(1)}_F + \mathbb{P}^{(1)}_M \ge 1-\delta_1, \label{eq: covertness criterion 1}\\
    &\lim\limits_{n\to\infty} \mathbb{P}_F^{(2)} + \mathbb{P}_M^{(2)} \ge 1- \delta_2, \label{eq: covertness criterion 2}\\
    &\lim\limits_{n\to\infty} P_\textnormal{error}^{(n)} = 0, \label{eq: covertness criterion 3}
    \end{align}
    where
    \begin{align}
        \mathbb{P}^{(1)}_F &\triangleq \Pr(\widehat{H}_e=H_1|H_0)\label{eq: P_F^1 definition}\\
        \mathbb{P}^{(1)}_M &\triangleq \Pr(\widehat{H}_e=H_0|H_1)\label{eq: P_M^1 definition}\\
        \mathbb{P}_F^{(2)}&\triangleq \Pr(\widehat{H}_c = \widetilde{H}_1|\widetilde{H}_0,H_1,\breve{H}_e=H_0)\label{eq: P_F^2 definition}\\
        \mathbb{P}_M^{(2)}&\triangleq \Pr(\widehat{H}_c=\widetilde{H}_0|\widetilde{H}_1,H_1,\breve{H}_e=H_0).\label{eq: P_M^2 definition}
    \end{align}
    with
    \begin{align}
        \breve{H}_e &= \begin{cases}
        H_0, & \textnormal{if } \mathbb{P}^{(1)}_F + \mathbb{P}^{(1)}_M > 1-\delta_1\\
        \hat{H}_e, & \textnormal{otherwise}
        \end{cases}\label{eq: willie's assumption}.
    \end{align}
    Here $\widehat{H}_e$ and $\widehat{H}_c$ denote Willie's decision in the channel estimation and the communication phases, respectively. $\breve{H}_e$ denotes the hypothesis in the channel estimation phase based on which Willie will conduct his test in the communication phase. $P_\text{error}^{(n)}$ denotes the probability that Willie cannot decode $\bm{x}_A$.
    We require that $\delta_1=\Theta_L(1)$ and $\delta_2=\Theta_n(1)$ be small but non-vanishing constants.
\end{defn}

Our main objective is to find whether for given $(\delta_1,\delta_2,\Lambda_A,R_A)$, Tom can drive the system to the linear regime, i.e., communicate with Eve covertly at a non-zero rate by a proper choice of $\epsilon$ and $\Lambda_T$, and, if so, to study this covert rate.

\section{Main Results}
\label{sec: main result}
We now state and prove our main results regarding the achievability of positive covert rate. We first consider a positive detection budget, i.e. $\delta_1,\delta_2>0$. Our main result in Theorem~\ref{thm: main result 1} is an achievable set of covert rates. 
\begin{thm}{\textbf{(Achievable Covert Rate when $\delta_1>0$)}} \label{thm: main result 1} Consider a \emph{detection budget} $(\delta_1,\delta_2)$ with $\delta_1\in(0,1)$ and $\delta_2\in(0,1)$, and Alice's transmit power and rate pair as $(\Lambda_A,R_A)$. Assume Tom's scaling parameter $\epsilon$ and transmit power $\Lambda_T$ satisfy 
\begin{align}
    \epsilon &\le \frac{\delta_1}{\sqrt{2}}, \label{eq: main result condition 1}\\
    \tau(\epsilon) &< \epsilon^2 \alpha_W^2 |h_W|^2 \Lambda_A + \sigma_W^2, \label{eq: main result condition 2}\\
    R_A&\le \log_2 (1+\gamma_W), \label{eq: main result condition 3}
\end{align}
where 
\begin{align}
    \tau(\epsilon) &\triangleq (1+\epsilon)^2\alpha_W^2|h_W|^2 \Lambda_T \frac{\exp\left(\frac{(1+\epsilon)^2\alpha_W^2|h_W|^2\Lambda_T}{\sigma_W^2}\right)}{\exp\left(\frac{(1+\epsilon)^2\alpha_W^2|h_W|^2\Lambda_T}{\sigma_W^2}\right)-1},\label{eq: asymptotic threshold}\\
    \gamma_W &= \frac{\alpha_W^2|h_W|^2 \Lambda_A}{\epsilon^2\alpha_W^2|h_W|^2 \Lambda_A+\alpha_W^2 |h_W|^2 \Lambda_T+\sigma_W^2}. \label{eq: willie's sinr}
\end{align}
Then, Tom can communicate with Eve covertly at any rate $R_T$ satisfying
\begin{align}
    R_T &\le \log_2\left(1+\frac{\alpha_E^2 |h_E|^2 \Lambda_T}{\alpha_E^2 |h_E|^2 \Lambda_A+\sigma_E^2}\right). \label{eq: covert rate without interference cancellation}
\end{align}
Additionally, if 
\begin{align}
    R_A&\le \log_2 \left(1+\frac{\alpha_E^2 |h_E|^2 \Lambda_A}{\alpha_E^2 |h_E|^2 \Lambda_T + \sigma_E^2}\right) \label{eq: main result condition 4}
\end{align}
then, Tom's rate $R_T$ can be improved to
\begin{align}
    R_T &\le \log_2\left(1+\frac{\alpha_E^2 |h_E|^2 \Lambda_T}{\sigma_E^2}\right). \label{eq: covert rate with interference cancellation}
\end{align}
\end{thm}

Theorem~\ref{thm: main result 1} states that as long as $\delta_1>0$, for any $\delta_2\in(0,1)$ Tom can communicate with Eve at a positive covert rate, effectively operating in the linear covert regime. 

As discussed in detail in Sections~\ref{subsec: channel estimation phase} through \ref{subsec: willie's SINR degradation}, the conditions in Theorem~\ref{thm: main result 1} have the following interpretation: Eq.~\eqref{eq: main result condition 1} corresponds to Tom satisfying the covertness criterion Eq.~\eqref{eq: covertness criterion 1} in the channel estimation phase. Subsequently, if Eq.~\eqref{eq: main result condition 2} is satisfied, then Willie's test decides in favor of $\widetilde{H}_1$ independent of $\bm{y}^{\textnormal{comm}}$, leading to a blind test, ensuring Eq.~\eqref{eq: covertness criterion 2}. Eq.~\eqref{eq: main result condition 3} ensures that Tom's actions do not disrupt communication over the legitimate link, satisfying Eq.~\eqref{eq: covertness criterion 3}. Note that to achieve $R_T$ in Eq.~\eqref{eq: covert rate without interference cancellation}, Eve treats Alice's signal as noise. Finally, the additional constraint Eq.~\eqref{eq: main result condition 4} allows Eve to perform interference cancellation, by which she decodes and cancels the legitimate signal $\bm{x}_A$ by first treating $\bm{x}_T$ as noise, leading to the covert rate $R_T$ in Eq.~\eqref{eq: covert rate with interference cancellation}.

\begin{figure}[t]
\centerline{
\iftoggle{shortversion}{
\includegraphics[width=0.4\textwidth]{Figures/achievablerate.eps}
}
{
\includegraphics[width=0.5\textwidth,trim={0cm 6cm 1cm 6cm},clip]{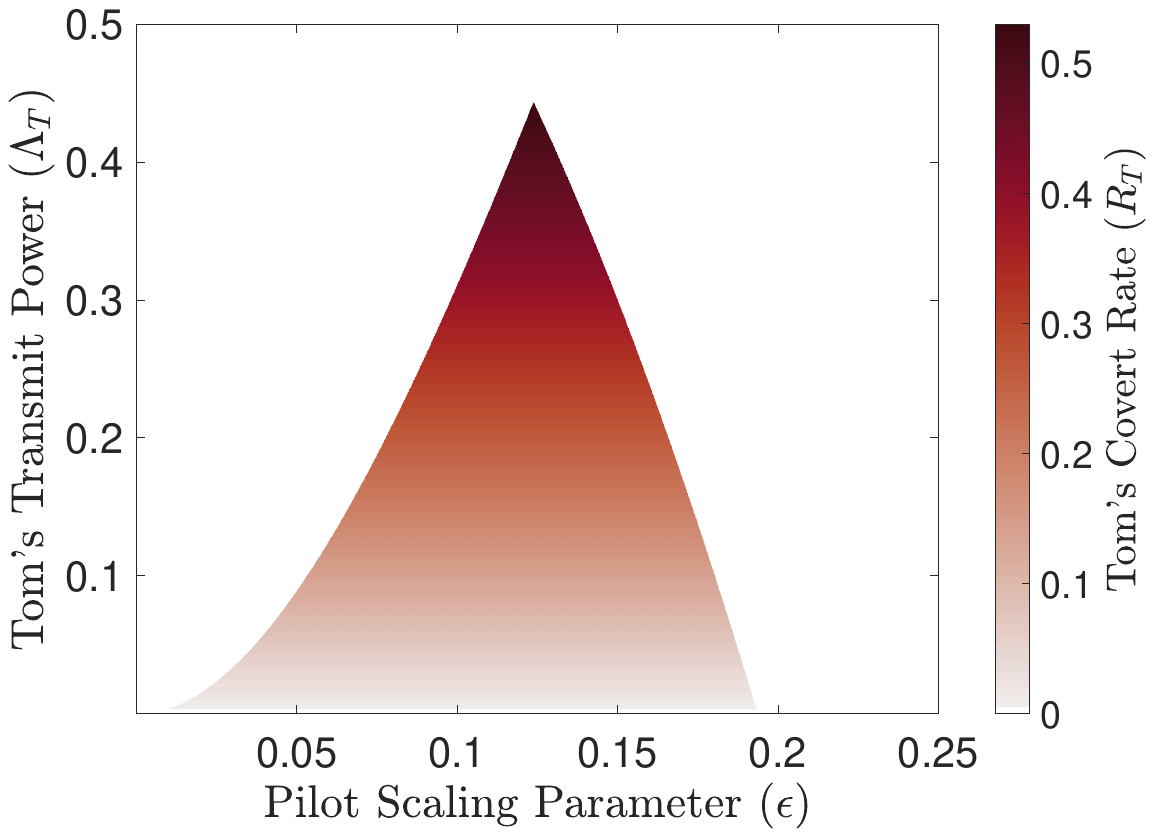}
}
}
\caption{A heatmap demonstrating the relationship between Tom's pilot scaling parameter $\epsilon$, his transmit power $\Lambda_T$, and the achievable covert rate $R_T$ given in Theorem~\ref{thm: main result 1}, where colors indicate the values of $R_T$ across the range of $\epsilon$ and $\Lambda_T$ when Eve can perform interference cancellation (See Eq.~\eqref{eq: main result condition 4}). Here, $\alpha_W^2=\alpha_E^2=0.1$, $|h_W|^2=|h_E|^2=1$, $\sigma_W^2=\sigma_E^2=0.1$, $\delta_1=\nicefrac{1}{\sqrt{10}}$, and $\Lambda_A = 20$ where Alice transmits at ($\approx$ 3.5 bpcu) $80\%$ of her capacity ($\approx 4.4$ bpcu) to Willie (See Eq.~\eqref{eq: alice's rate}).}
\label{fig: achievable rate}
\end{figure}

Figure~\ref{fig: achievable rate} illustrates the $(\epsilon,\Lambda_T)$ pairs satisfying Theorem~\ref{thm: main result 1} and the corresponding achievable positive covert rates $R_T$ for a given configuration. A key observation is that the covert rate is not necessarily maximized when Tom maximizes $\epsilon$ and the channel estimation error of Willie. This happens because Willie's signal-to-interference-plus-noise ratio (SINR) degrades as a result of both mismatched decoding due to his imperfect channel estimation.

Next, we consider a zero detection budget in the channel estimation phase. Our main result regarding the achievable covert rate when $\delta_1=0$ is presented in Theorem~\ref{thm: main result 2} below. 

\begin{thm}{\textbf{(Achievable Covert Rate when $\delta_1=0$)}}\label{thm: main result 2}
    When $\delta_1=0$, Tom can transmit covertly
    if and only if
    \begin{align}
        R_T = \mathcal{O}(n^{-\nicefrac{1}{2}})
    \end{align}
    for any $\delta_2\in(0,1)$.
\end{thm}
Theorem~\ref{thm: main result 2} states that when $\delta_1=0$, Tom's scaling parameter $\epsilon$ needs to vanish with $L$ and hence Tom cannot conduct an effective pilot scaling attack and in turn, he needs to obey the square-root law~\cite{Bash2013}.

The rest of this section details the proof of Theorems~\ref{thm: main result 1}~and~\ref{thm: main result 2}. First, we focus on the channel estimation phase in Section~\ref{subsec: channel estimation phase} and we investigate the covertness of Tom's pilot scaling attack as well as its impact on Willie's channel estimate. Next, we consider the communication phase in Section~\ref{subsec: willie's optimal threshold} and discuss Willie's detection strategy and his optimal threshold. In Section~\ref{subsec: positive power}, we derive sufficient and necessary conditions under which Tom could transmit at a non-vanishing power $\Lambda_T$ while remaining covert. In Section~\ref{subsec: willie's SINR degradation}, we investigate the SINR deterioration at Willie due to Tom's actions. Finally, in Sections~\ref{subsec: proof of main result 1}~and~\ref{subsec: proof of main result 2} we present the complete proofs of Theorems~\ref{thm: main result 1}~and~\ref{thm: main result 2}.

\subsection{Covert Pilot Scaling \& Willie's Channel Estimation Error}
\label{subsec: channel estimation phase}
We start by deriving sufficient conditions for Tom's scaling parameter $\epsilon$ such that the covertness criterion Eq.~\eqref{eq: covertness criterion 1} in Definition~\ref{defn: covertness criterion} is met.

\begin{lem}{\textbf{(Covert Pilot Scaling)}} \label{lem: covert pilot scaling} For any $\epsilon\le \nicefrac{\delta_1}{\sqrt{2}}$, Tom's pilot scaling attack remains covert.
\end{lem}
\iftoggle{shortversion}{
\begin{sketchproof}
As is customary, we start by observing
\begin{align}
    \mathbb{P}^{(1)}_F + \mathbb{P}^{(1)}_M &\ge 1-\sqrt{\mathbb{D}(\mathbb{P}_1\|\mathbb{P}_0)}
\end{align}
where $\mathbb{D}(\mathbb{P}_1\|\mathbb{P}_0)$ denotes the Kullback–Leibler divergence  between the alternative $\mathbb{P}_1$ and the null $\mathbb{P}_0$ distributions induced by $\bm{y}^{\textnormal{est}}_W$ under both hypotheses. We obtain
\begin{align}
    \lim\limits_{L\to\infty} \mathbb{D}(\mathbb{P}_1\|\mathbb{P}_0) &= 2 \log(1+\epsilon)-1+(1+\epsilon)^{-2}
\end{align}
and argue that $\lim\limits_{L\to\infty} \mathbb{D}(\mathbb{P}_1\|\mathbb{P}_0)\le 2\epsilon^2$.
\end{sketchproof}
}
{
\begin{proof}
    See Appendix~\ref{proof: covert pilot scaling}.
\end{proof}
}

For $\epsilon\le \nicefrac{\delta_1}{\sqrt{2}}$, $\breve{H}_e=H_0$ and Willie estimates $h_W$ assuming the null hypothesis $H_0$ is true. As customary, we assume Willie uses the minimum mean square error (MMSE) estimator $\hat{h}_W$ of $h_W$. 

\begin{prop}{\textbf{(Willie's Estimate)}}\label{prop: mmse estimate}
For $\epsilon\le \nicefrac{\delta_1}{\sqrt{2}}$, Willie's estimate $\hat{h}_W\triangleq\lim\limits_{L\to\infty}\hat{h}^{\textnormal{mmse}}_W$ under $H_0$ and $H_1$ is given in
\begin{align}
    \begin{array}{ll}
       H_0:  & \hat{h}_W=h_W , \\
       H_1:  & \hat{h}_W=(1+\epsilon)h_W.
    \end{array}
    \label{eq: willie's estimate under hypotheses}
\end{align}
\end{prop}

\iftoggle{shortversion}{
}
{
\begin{proof}
    See Appendix~\ref{proof: mmse estimate}.
\end{proof}
}

\subsection{Willie's Detection in the Communication Phase}
\label{subsec: willie's optimal threshold}
In this subsection, assuming Tom's pilot scaling attack remains covert (Eq.~\eqref{eq: covertness criterion 1}), i.e., $\breve{H}_e=H_0$, we discuss Willie's detection strategy for Tom in the communication phase. Throughout, we assume Willie can decode $\bm{x}_A$ as per Eq.~\eqref{eq: covertness criterion 3}. 

Note that, from Eq.~\eqref{eq: received signal general model} and Eq.~\eqref{eq: transmitted signal}, the received signal at Willie is given by
\begin{align}
\begin{array}{ll}
     \widetilde{H}_0:& \bm{y}^{\textnormal{comm}}_W = \alpha_W \: h_W \:{\bm{x}_A} + \bm{z}^{\textnormal{comm}}_W,\\
     \widetilde{H}_1:& \bm{y}^{\textnormal{comm}}_W = \alpha_W \: h_W \:{\bm{x}_A} + \alpha_W \: h_W \:{\bm{x}_T} + \bm{z}^{\textnormal{comm}}_W.
\end{array}
\label{eq: received signal}
\end{align}

In Lemma~\ref{lem: willie's optimal test} below, we show that Willie's optimal detection strategy is to adopt a radiometer, similar to \cite{sobers2017covert,ta2019covert}.

\begin{lem}{\textbf{(Willie's Optimal Detection Strategy)}}\label{lem: willie's optimal test}
    Given that Willie can decode $\bm{x}_A$ correctly, under $H_0$ Willie's most powerful test and the optimal threshold in the communication phase (minimizing the sum of the false alarm and missed detection probabilities) is given by
    \begin{align}
    T(\bm{y}^{\textnormal{comm}}_W) & \triangleq \frac{1}{n}\|\bm{y}^{\textnormal{comm}}_W-\alpha_W \hat{h}_W\bm{x}_A\|_2^2 \underset{\widetilde{H}_0}{\overset{\widetilde{H}_1}{\gtrless}}\tau^\dagger\label{eq: willie's test}
\end{align}
where
\begin{align}
    \tau^\dagger \triangleq \alpha_W^2|\hat{h}_W|^2 \Lambda_T \frac{\exp\left(\frac{n}{n-1}\frac{\alpha_W^2|\hat{h}_W|^2\Lambda_T}{\sigma_W^2}\right)}{\exp\left(\frac{n}{n-1}\frac{\alpha_W^2|\hat{h}_W|^2\Lambda_T}{\sigma_W^2}\right)-1} \label{eq: willie's threshold}
\end{align}
and $\hat{h}_W$ is Willie's estimate of $h_W$.
\end{lem}
\iftoggle{shortversion}{
\begin{sketchproof}
The test statistic $T(\bm{y}^{\textnormal{comm}}_W)$ is found by computing and simplifying the log-likelihood ratio based on $\bm{y}^{\textnormal{comm}}_W$. Then, the optimal threshold is given by
\begin{align}
    \tau^\dagger &\triangleq \argmin\limits_{\tau>0} \mathbb{P}^\dagger_F+\mathbb{P}^\dagger_M
\end{align}
where
\begin{align}
    \mathbb{P}^\dagger_F(\tau) &\triangleq \Pr(T(\bm{y}^{\textnormal{comm}}_W)>\tau|\widetilde{H}_0,H_0)\label{eq: P_F_dagger}\\
    \mathbb{P}^\dagger_M(\tau) &\triangleq \Pr(T(\bm{y}^{\textnormal{comm}}_W)<\tau|\widetilde{H}_1,H_0)\label{eq: P_M_dagger}
\end{align}
Solving $\smash{\frac{\partial (\mathbb{P}^\dagger_F(\tau)+\mathbb{P}^\dagger_M(\tau))}{\partial \tau}\Big|_{\tau=\tau^\dagger}} = 0$,
we obtain Eq.~\eqref{eq: willie's threshold}.
\end{sketchproof}
}
{
\begin{proof}
    See Appendix~\ref{proof: willie's optimal test}.
\end{proof}
}

Observe that a key difference between our system model and those of \cite{sobers2017covert,ta2019covert} is the existence of the legitimate signaling by Alice. Furthermore, Lemma~\ref{lem: willie's optimal test} suggests that before using the radiometer, Willie decodes and cancels the legitimate signal $\bm{x}_A$. In fact, under $H_0$, Willie's channel estimate $\hat{h}_W$ is perfect and hence our test statistic $T(\bm{y}^{\textnormal{comm}}_W)$ corresponds to that of \cite{sobers2017covert,ta2019covert}.

Since Willie is unaware of pilot scaling by $\epsilon$ and in turn believes his cancellation of $\mathbf{x}_A$ has been perfect, his test threshold $\tau^\dagger$ is independent of $\Lambda_A$. Furthermore, one can easily verify that $\tau^\dagger$ is increasing in $n$.

\subsection{Covertly Transmitting At a Positive Power}
\label{subsec: positive power}
When Tom performs a pilot scaling attack and remains covert, i.e., $\breve{H}_e=H_0$ while $H_1$ is true, using Proposition~\ref{prop: mmse estimate}, Willie's test statistic $T(\bm{y}^{\textnormal{comm}}_W)$ becomes
\begin{align}
\hspace{-1em}
\begin{array}{ll}
     \widetilde{H}_0: &\hspace{-0.5em} T(\bm{y}_W^{\textnormal{comm}})=\epsilon^2 \alpha_W^2 |h_W|^2 \Lambda_A + \Lambda_Z  \\
     \widetilde{H}_1: &\hspace{-0.5em}  T(\bm{y}_W^{\textnormal{comm}})=\epsilon^2 \alpha_W^2 |h_W|^2 \Lambda_A +\alpha_W^2 |h_W|^2 \Lambda_T+\Lambda_Z
\end{array}
\label{eq: test statistic after imperfect cancellation}
\end{align}
where $\Lambda_Z\triangleq\frac{1}{n}\|\bm{z}^{\textnormal{comm}}_W\|_2^2$.
Because of the imperfect cancellation of the legitimate signal $\bm{x}_A$, there is a residual term depending on $\epsilon$ and $\Lambda_A$ under both hypotheses in Eq.~\eqref{eq: test statistic after imperfect cancellation} (See also Eq.~\eqref{eq: willie's estimate under hypotheses}). 

Under $H_1$, we have $\tau(\epsilon)=\lim\limits_{n\to\infty} \tau^\dagger$, where $\tau(\epsilon)$ is given in Eq.~\eqref{eq: asymptotic threshold}. When $\breve{H}_e=H_0$, for sufficiently large $n$, Willie's test will be
\begin{align}
    T(\bm{y}^{\textnormal{comm}}_W) & \underset{\widetilde{H}_0}{\overset{\widetilde{H}_1}{\gtrless}}\tau(\epsilon).
\end{align}
Now we analyze the performance of Willie's test under $H_1$ in terms of $\mathbb{P}_F^{(2)}$ and $\mathbb{P}_M^{(2)}$.

\begin{lem}{\textbf{(Covert Communications with Pilot Scaling)}}\label{lem: covert transmission}
    As long as Willie's optimal threshold satisfies
    \begin{align}
        \tau(\epsilon) < \epsilon^2 \alpha_W^2 |h_W|^2 \Lambda_A + \sigma_W^2, \label{eq: covertness threshold 1}
    \end{align}
    or
    \begin{align}
        \tau(\epsilon) > \alpha_W^2 |h_W|^2 (\epsilon^2  \Lambda_A+ \Lambda_T) + \sigma_W^2, \label{eq: covertness threshold 2}
    \end{align}
    we have $\lim\limits_{n\to\infty} \mathbb{P}^{(2)}_F + \mathbb{P}^{(2)}_M = 1$.

    Conversely, if 
    \begin{align}
        \hspace{-0.5em}\epsilon^2 \alpha_W^2 |h_W|^2 \Lambda_A + \sigma_W^2&<\tau(\epsilon)< \alpha_W^2 |h_W|^2 (\epsilon^2 \Lambda_A + \Lambda_T) + \sigma_W^2,\label{eq: lemma 3 converse}
    \end{align}
    we have $\lim\limits_{n\to\infty} \mathbb{P}^{(2)}_F + \mathbb{P}^{(2)}_M = 0$.
\end{lem}
\iftoggle{shortversion}{
\begin{sketchproof}
    The proof follows from the observation that $\frac{2}{\sigma_W^2}\frac{1}{n}\|\bm{z}^{\textnormal{comm}}_W\|_2^2\sim\chi^2(2n)$ and the subsequent use of the tail bounds~\cite[Lemma 1]{laurent2000adaptive} for the $\chi^2$-distribution on $T(\bm{y}_W^{\textnormal{comm}})$ (See Eq.~\eqref{eq: test statistic after imperfect cancellation}) in order to bound $\mathbb{P}^{(2)}_F$, $\mathbb{P}^{(2)}_M$, $1-\mathbb{P}^{(2)}_F$, or $1-\mathbb{P}^{(2)}_M$ from above, depending on the assumptions on $\tau(\epsilon)$.
\end{sketchproof}
}
{
\begin{proof}
    See Appendix~\ref{proof: covert transmission}.
\end{proof}
}
Note that when $H_1$ is true and $\breve{H}_e=H_0$, as $n\to\infty$ we have
\begin{align}
\begin{array}{ll}
     \widetilde{H}_0:& T(\bm{y}^{\textnormal{comm}}_W)\overset{p}{\to}\epsilon^2 \alpha_W^2 |h_W|^2 \Lambda_A + \sigma_W^2 \\
     \widetilde{H}_1:& T(\bm{y}^{\textnormal{comm}}_W)\overset{p}{\to}\alpha_W^2 |h_W|^2 (\epsilon^2 \Lambda_A + \Lambda_T) + \sigma_W^2
\end{array}
\end{align}
Hence, Lemma~\ref{lem: covert transmission} states that Tom can covertly transmit to Eve only if he can adjust $\Lambda_T$ and $\epsilon$ such that Willie sets his test threshold $\tau(\epsilon)$ either below the limit values of $T(\bm{y}^{\textnormal{comm}}_W)$ under both $\widetilde{H}_0$ and $\widetilde{H}_1$, or above both of these limit values.

Furthermore, Lemma~\ref{lem: covert transmission} implies that given $\epsilon>0$, as Alice's transmit power $\Lambda_A$ increases, Tom's chances at covertness improve as the residual term in Eq.~\eqref{eq: test statistic after imperfect cancellation} due to imperfect channel estimation also increases with $\Lambda_A$. 

Recall that the optimal threshold $\tau(\epsilon)$ is an increasing function of both $\epsilon$ and $\Lambda_T$, and the RHS of Eq.~\eqref{eq: covertness threshold 1} is a function of $\epsilon$ and $\Lambda_A$. Hence, there exists $\Lambda^\ast_T=\Theta_n(1)$ such that
\begin{align}
    \tau(\epsilon)|_{\Lambda_T=\Lambda^\ast_T} = \epsilon^2 \alpha_W^2 |h_W|^2 \Lambda_A + \sigma_W^2.
\end{align}
Therefore, Lemma~\ref{lem: covert transmission} implies that based on $\Lambda_A$ and $\epsilon$, Willie can covertly transmit at a non-vanishing power $\Lambda_T < \Lambda^\ast_T$.

Conversely, we argue that when there is no pilot scaling attack, i.e., when $\epsilon=0$, Tom cannot transmit at a non-vanishing power $\Lambda_T$ covertly. More specifically, in Lemma~\ref{lem: square root law} below we provide a sufficient and necessary condition on $\Lambda_T$ for Tom to remain covert.

\begin{lem}{\textbf{(Covert Communications with No Pilot Scaling)}}\label{lem: square root law}
    When $\epsilon=0$, Tom can only transmit covertly when he transmits at a power $\Lambda_T = \mathcal{O}(n^{-\nicefrac{1}{2}})$.
\end{lem}

\iftoggle{shortversion}{
\begin{sketchproof}
    The proof starts with proving that when $\epsilon=0$, $\tau(\epsilon)$ satisfies Eq.~\eqref{eq: lemma 3 converse} for any $\Lambda_T=\Theta_n(1)$. Hence, by Lemma~\ref{lem: covert transmission}, Tom cannot transmit covertly at any non-vanishing power $\Lambda_T=\Theta_n(1)$. Therefore, to remain covert, Tom needs to transmit at a power $\Lambda_T=o_n(1)$. Next, we assume $\Lambda_T=\mathcal{O}(n^{-\nicefrac{1}{2}})$ and prove the achievability part via an upper bound on $1-\mathbb{P}^{(2)}_F-\mathbb{P}^{(2)}_M$ by a right Riemann sum and further bounding this Riemann sum via Stirling's approximation~\cite[Chapter 3.2]{cormen2022introduction}. Finally, for the converse result, we assume $\Lambda_T=\omega(n^{-\nicefrac{1}{2}})$ and use tail bounds for $\chi^2$-distribution to show that $\mathbb{P}^{(2)}_F+\mathbb{P}^{(2)}_M\to 0$ as $n\to\infty$. 
\end{sketchproof}
}
{
\begin{proof}
    See Appendix~\ref{proof: square root law}.
\end{proof}
}

\subsection{Willie's SINR Degradation}
\label{subsec: willie's SINR degradation}
We stress that when Tom conducts a pilot scaling attack with some $\epsilon>0$ and subsequently transmits at a non-vanishing power $\Lambda_T$, Willie's SINR deteriorates. More formally, Willie's SINR will become
\begin{align}
    \gamma_W &= \frac{\alpha_W^2|h_W|^2 \Lambda_A}{\epsilon^2\alpha_W^2|h_W|^2 \Lambda_A+\alpha_W^2 |h_W|^2 \Lambda_T+\sigma_W^2}. \label{eq: willie's SINR}
\end{align}

Note that in Eq.~\eqref{eq: willie's SINR} the first interference term stems from the mismatched decoding caused by Tom's pilot scaling attack and in turn Willie's imperfect channel estimation, while the second is caused by Tom's transmission.

Furthermore, if $\epsilon$ and $\Lambda_T$ are large enough such that 
\begin{align}
    R_A > \log_2 (1+\gamma_W),
\end{align}
Willie will start having decoding errors. Since this unexpected decoding error will imply the existence of a rogue communication, Tom needs to avoid it. 

Observe that only when $\Lambda_T$ is comparable to or much larger than $\Lambda_A$, the constraint Eq.~\eqref{eq: covertness threshold 2} in Lemma~\ref{lem: covert transmission} can be satisfied. Since this will disrupt the legitimate Alice-Willie link and in turn be detected, we only focus on satisfying Eq.~\eqref{eq: covertness threshold 1}.

\subsection{Proof of Theorem~\ref{thm: main result 1}}
\label{subsec: proof of main result 1}
We now prove Theorem~\ref{thm: main result 1}. Observe that Lemma~\ref{lem: covert pilot scaling} states that as long as Eq.~\eqref{eq: main result condition 1} is satisfied, the first covertness criterion Eq.~\eqref{eq: covertness criterion 1} is satisfied. Hence given Eq.~\eqref{eq: main result condition 1}, Willie performs channel estimation based on $\breve{H}_e=H_0$ (See Proposition~\ref{prop: mmse estimate}).

Since Tom is successful in the channel estimation phase, Willie is unaware of the pilot scaling attack and conducts his optimal detection strategy in the communication phase as described in Proposition~\ref{lem: willie's optimal test}. Noticing that $\tau(\epsilon) = \lim\limits_{n\to\infty} \tau^\dagger$, with $\tau(\epsilon)$ given Eq.~\eqref{eq: asymptotic threshold}, Lemma~\ref{lem: covert transmission} states that the second covertness criterion Eq.~\eqref{eq: covertness criterion 2} is satisfied. 

Finally, as described in Section~\ref{subsec: willie's SINR degradation}, Eq.~\eqref{eq: main result condition 3} ensures that the Alice-Willie link is not disrupted, hence ensuring the final covertness criterion (See Eq.~\eqref{eq: covertness criterion 3}.) Therefore, given Eq.~\eqref{eq: main result condition 1}-\eqref{eq: main result condition 3}, Tom communicates covertly with Eve who treats $\bm{x}_A$ as noise, yielding the achievable rate of Eq.~\eqref{eq: covert rate without interference cancellation}. 

As stated in Section~\ref{sec: main result}, if Eq.~\eqref{eq: main result condition 4} is satisfied, Eve first successfully decodes and cancels $\bm{x}_A$, treating $\bm{x}_T$ as noise. Note that Eve is aware of Tom's pilot scaling attack and hence her channel estimation and her subsequent cancellation of $\bm{x}_A$ are perfect. Thus, we obtain the improved rate given in Eq.~\eqref{eq: covert rate with interference cancellation}. \qed

\subsection{Proof of Theorem~\ref{thm: main result 2}}
\label{subsec: proof of main result 2}
Next, we prove Theorem~\ref{thm: main result 2}. Begin by observing that, $\delta_1=0$ necessitates $\epsilon=o_L(1)$. Hence by Proposition~\ref{prop: mmse estimate}, $\hat{h}_W = h_W$
under both $H_0$ and $H_1$. In other words, for any $\epsilon=o_L(1)$, Tom's pilot scaling attack has no impact on Willie's channel estimation process. Thus, we can only focus on the $\epsilon=0$ case.

Next, note that when $\epsilon=0$ by Lemma~\ref{lem: square root law}, Tom can communicate covertly with Eve only when $\Lambda_T=\mathcal{O}(n^{-\nicefrac{1}{2}})$. 

Finally, by performing the MacLaurin series expansion of the subsequent achievable rate with respect to $\Lambda_T$, we conclude that when $\epsilon=o_L(1)$, Tom can covertly communicate with Eve at a rate $R_T$ if and only if $R_T=\mathcal{O}(n^{-\nicefrac{1}{2}})$.\qed

\section{Conclusion}
\label{sec: conclusion}
We have investigated a covert communications scenario in which a hardware Trojan carries out a pilot scaling attack to degrade the channel estimate of legitimate parties and subsequently reduces their ability to detect the Trojan's communication. We have showed that for any positive pilot detection budget, the Trojan can effectively drive the system to the linear regime, allowing non-zero covert communication rates. Conversely, we have shown that in the zero pilot detection budget case, the Trojan loses its ability to covertly and effectively corrupt the channel estimation process and in turn has to obey the square root law. Overall, our findings suggest that effective strategies against hardware Trojans also need to take into account the channel estimation phase.


\typeout{}
\bibliographystyle{IEEEtran}
\bibliography{references}

\iftoggle{shortversion}{}
{
\appendix
\subsection{Proof of Lemma~\ref{lem: covert pilot scaling}}\label{proof: covert pilot scaling}

From Eq.~\eqref{eq: received signal general model} and \eqref{eq: transmitted pilot sequence}, the received pilot sequence at Willie is given by 
\begin{align}
\begin{array}{ll}
     H_0:& \bm{y}^{\textnormal{est}}_W = \alpha_W  h_W {\bm{s}_A} + \bm{z}^{\textnormal{est}}_W \\
     H_1:& \bm{y}^{\textnormal{est}}_W =\alpha_W  h_W (1+\epsilon){\bm{s}_A} + \bm{z}^{\textnormal{est}}_W
\end{array}
\label{eq: received pilot sequence}
\end{align}
Next, we argue that
\begin{align}
\begin{array}{ll}
     H_0:& \bm{y}^{\textnormal{est}}_W\sim \mathcal{CN}(\bm{0},\bm{\Sigma}_0)  \\
     H_1:& \bm{y}^{\textnormal{est}}_W\sim \mathcal{CN}(\bm{0},\bm{\Sigma}_1)
\end{array}
\end{align}
where 
\begin{align}
    \bm{\Sigma}_0 &= \alpha_W^2 \sigma_H^2 \bm{s}_A \bm{s}_A^H + \sigma_W^2 \bm{I}\\
    \bm{\Sigma}_1 &= \alpha_W^2 \sigma_H^2 (1+\epsilon)^2 \bm{s}_A \bm{s}_A^H + \sigma_W^2 \bm{I}
\end{align}

Note that from~\cite[Theorem 13.1.1]{lehmann1986testing} and \cite[Lemma 11.6.1]{cover2006elements}, we obtain
\begin{align}
    \mathbb{P}^{(1)}_F + \mathbb{P}^{(1)}_M &\ge 1-\sqrt{\mathbb{D}(\mathbb{P}_1\|\mathbb{P}_0)}
\end{align}
where $\mathbb{D}(\mathbb{P}_1\|\mathbb{P}_0)$ denotes the Kullback–Leibler divergence  between the alternative $\mathbb{P}_1$ and the null $\mathbb{P}_0$ distributions induced by $\bm{y}^{\textnormal{est}}_W$ under both hypotheses. Furthermore $\mathbb{D}(\mathbb{P}_1\|\mathbb{P}_0)$ can be computed as~\cite{duchi2007derivations}
\begin{align}
   \mathbb{D}(\mathbb{P}_1\|\mathbb{P}_0)= -\log |\bm{\Sigma}_1^{-1}\bm{\Sigma}_0| -L + \text{tr}(\bm{\Sigma}_1^{-1}\bm{\Sigma}_0) \label{eq: KL term}
\end{align}
Note that Tom's goal is to stay covert by keeping $\bm{\Sigma}_1\approx \bm{\Sigma}_0$ and in turn $\bm{\Sigma}_1^{-1}\bm{\Sigma}_0\approx \bm{I}$ and $\mathbb{D}(\mathbb{P}_1\|\mathbb{P}_0)\approx 0$.

Now, since $\bm{s}_A \bm{s}_A^H$ is rank one, we can compute $\bm{\Sigma}_1^{-1}$ as~\cite{miller1981inverse}
\begin{align}
    \bm{\Sigma}_1^{-1} &= \frac{1}{\sigma^2_W}\left(\bm{I}-\frac{\alpha_W^2 \sigma_H^2 (1+\epsilon)^2}{\sigma_W^2 + \alpha_W^2 \sigma_H^2 (1+\epsilon)^2 \|\bm{s}_A\|_2^2} \bm{s}_A \bm{s}_A^H\right) \label{eq: sigma1 inverse}
\end{align}
This, in turn, leads to
\begin{align}
    \bm{\Sigma}_1^{-1}\bm{\Sigma}_0 = \bm{I} &+ \frac{\alpha_W^2 \sigma_H^2}{\sigma_W^2} \bm{s}_A \bm{s}_A^H\notag\\
    &- \frac{\alpha_W^2 \sigma_H^2 (1+\epsilon)^2}{\sigma_W^2 + \alpha_W^2 \sigma_H^2 (1+\epsilon)^2 \|\bm{s}_A\|_2^2} \bm{s}_A \bm{s}_A^H\notag\\
    &- \frac{\frac{\alpha_W^4 \sigma_H^4}{\sigma_W^2} (1+\epsilon)^2\|\bm{s}_A\|_2^2}{\sigma_W^2 + \alpha_W^2 \sigma_H^2 (1+\epsilon)^2 \|\bm{s}_A\|_2^2} \bm{s}_A \bm{s}_A^H \label{eq: matrixprod}
\end{align}


From Eq.~\eqref{eq: matrixprod} and the fact that for any rank-one matrix $\bm{A}$
\begin{align}
    |\bm{A}+\bm{I}| &= 1+ \text{tr}(\bm{A}),
\end{align}
we get
\begin{align}
    |\bm{\Sigma}_1^{-1}\bm{\Sigma}_0| =1 - \frac{\alpha_W^2 \sigma_H^2}{\sigma_W^2}   \frac{\epsilon (2+\epsilon)\|\bm{s}_A\|_2^2}{1+ \frac{\alpha_W^2\sigma_H^2}{\sigma_W^2} (1+\epsilon)^2 \|\bm{s}_{A}\|_2^2}\label{eq: determinant of matrixprod}\\
    \text{tr}(\bm{\Sigma}_1^{-1}\bm{\Sigma}_0) = L - \frac{\alpha_W^2 \sigma_H^2}{\sigma_W^2}   \frac{\epsilon (2+\epsilon)\|\bm{s}_A\|_2^2}{1+ \frac{\alpha_W^2\sigma_H^2}{\sigma_W^2} (1+\epsilon)^2 \|\bm{s}_{A}\|_2^2} \label{eq: trace of matrixprod}
\end{align}

Plugging Eq.~\eqref{eq: determinant of matrixprod} and Eq.~\eqref{eq: trace of matrixprod} into Eq.~\eqref{eq: KL term}, and taking $L\to \infty$, we get
\begin{align}
    \lim\limits_{L\to\infty} \mathbb{D}(\mathbb{P}_1\|\mathbb{P}_0) &= 2 \log(1+\epsilon)-1+(1+\epsilon)^{-2}
\end{align}
Letting
\begin{align}
    \kappa(\epsilon)&\triangleq 2 \log(1+\epsilon)-1+(1+\epsilon)^{-2} - 2\epsilon^2
\end{align}
 and observing that $\kappa$ satisfies $\kappa(0) = 0$ and $\kappa^\prime(\epsilon)\le 0$, ${\forall \epsilon >0}$ we conclude that
\begin{align}
    \lim\limits_{L\to\infty} \mathbb{D}(\mathbb{P}_1\|\mathbb{P}_0) &\le 2\epsilon^2.
\end{align}
Thus, as long as $\epsilon\le \nicefrac{\delta_1}{\sqrt{2}}$, Tom's pilot scaling attack remains covert.
\qed

\subsection{Proof of Proposition~\ref{prop: mmse estimate}}\label{proof: mmse estimate}
Since $h_W$ is Rayleigh, the MMSE is equal to the linear MMSE, and $\hat{h}^{mmse}_W$ is given by 
\begin{align}
    \hat{h}^{mmse}_W &= \bm{r}_{h_W \tilde{\bm{y}}} \bm{\Sigma}_0^{-1} \bm{y}^{\textnormal{est}}_W \label{eq: lmmse formula}
\end{align}
where
\begin{align}
    \tilde{\bm{y}} &\triangleq \alpha_W h_W \bm{s}_A + \bm{z}^{\textnormal{est}}_W\\
    \bm{r}_{h_W \tilde{\bm{y}}} &\triangleq \mathbb{E}[h_W \bm{y}^{\textnormal{est},0\: H}_{W}]
\end{align}
We compute $\bm{r}_{h_W \tilde{\bm{y}}}$ as
\begin{align}
    \bm{r}_{h_W \tilde{\bm{y}}} & = \mathbb{E}[h_W(\alpha_W h_W \bm{s}_A + \bm{z}^{\textnormal{est}}_W)^H]
    = \alpha_W \sigma_H^2 \bm{s}_A^H  \label{eq: hr correlation matrix}
\end{align}
Similar to Eq.~\eqref{eq: sigma1 inverse}, we can compute $\bm{\Sigma}_0^{-1}$ as 
\begin{align}
    \bm{\Sigma}_0^{-1} &= \frac{1}{\sigma^2_W}\left(\bm{I}-\frac{\alpha_W^2 \sigma_H^2}{\sigma_W^2 + \alpha_W^2 \sigma_H^2  \|\bm{s}_A\|_2^2} \bm{s}_A \bm{s}_A^H\right) \label{eq: sigma0 inverse}
\end{align}

Plugging Eq.~\eqref{eq: hr correlation matrix} and Eq.~\eqref{eq: sigma0 inverse} into Eq.~\eqref{eq: lmmse formula}, after some simplifications, we get
\begin{align}
    \hat{h}_W^{mmse} &= \frac{\alpha_W \sigma_H^2}{\sigma_W^2 + \alpha_W^2 \sigma_H^2 \|\bm{s}_A\|_2^2} \bm{s}_A^H \bm{y}^{\textnormal{est}}_W \label{eq: lmmse estimate}
\end{align}


\begin{figure*}[t] 

\begin{align}
\begin{array}{ll}
     H_0:& \hat{h}_W^{\textnormal{mmse}} = h_W - \frac{h_W}{1+\frac{\alpha_W^2 \sigma_H^2}{\sigma_W^2}\|\bm{s}_A\|_2^2} + \frac{\frac{\alpha_W \sigma_H^2}{\sigma_W^2}}{1+\frac{\alpha_W^2 \sigma_H^2}{\sigma_W^2}\|\bm{s}_A\|_2^2} \bm{s}_A^{H} \bm{z}^{\textnormal{est}}_W \\
     H_1:& \hat{h}_W^{\textnormal{mmse}} = h_W - \frac{h_W}{1+\frac{\alpha_W^2 \sigma_H^2}{\sigma_W^2}\|\bm{s}_A\|_2^2} + \frac{\frac{\alpha_W \sigma_H^2}{\sigma_W^2}}{1+\frac{\alpha_W^2 \sigma_H^2}{\sigma_W^2}\|\bm{s}_A\|_2^2} \bm{s}_A^{H} \bm{z}^{\textnormal{est}}_W \:\: + \: \epsilon \frac{\alpha^2\sigma_H^2 \|\bm{s}_A\|_2^2}{\sigma_W^2 + \alpha_W^2\sigma_H^2 \|\bm{s}_A\|_2^2} h_W
\end{array}
\label{eq: lmmse estimate cases}
 \customtag
\end{align}
\hrulefill
\end{figure*}

Combining Eq.~\eqref{eq: received pilot sequence} and Eq.~\eqref{eq: lmmse estimate}, we obtain Eq.~\eqref{eq: lmmse estimate cases}. Taking the limit as $L\to\infty$ and observing that $\|\bm{s}_A\|_2^2=\omega_L(1)$ concludes the proof.
\qed

\subsection{Proof of Lemma~\ref{lem: willie's optimal test}}\label{proof: willie's optimal test}
We first show that $T(\bm{y}_W^{\textnormal{comm}})$ is the right statistic for Willie. Note that, from the Neyman-Pearson Lemma~\cite[Theorem 11.7.1]{cover2006elements}, conditioned on $H_0$, Willie's optimal test is a likelihood-ratio test of the form
\begin{align}
    \ell(\bm{y}_W^{\textnormal{comm}}|H_0)&\triangleq \log\frac{f_{\bm{Y}^{\textnormal{comm}}_W|H_0,\widetilde{H}_1}(\bm{y}^{\textnormal{comm}}_W|H_0,\widetilde{H}_1)}{f_{\bm{Y}^{\textnormal{comm}}_W|H_0,\widetilde{H}_0}(\bm{y}^{\textnormal{comm}}_W|H_0,\widetilde{H}_0)}\underset{\widetilde{H}_0}{\overset{\widetilde{H}_1}{\gtrless}}\tau_{\text{LLR}}\label{eq: llr test}
\end{align}
for some log-likelihood ratio threshold $\tau_{\text{LLR}}$. 

Also note that since $\bm{x}_A$ is decoded by Willie, we get
\begin{align}
    \mathbb{E}[\bm{y}^{\textnormal{comm}}_W|H_0] &= \alpha_W \hat{h}_W \bm{x}_A
\end{align}
under both $\widetilde{H}_0$ and $\widetilde{H}_1$. Thus, we get
\begin{align}
    \ell(\bm{y}^{\textnormal{comm}}_W|H_0)&= -\frac{1}{2} \bm{v}^H (\tilde{\bm{\Sigma}}_1^{-1}-\tilde{\bm{\Sigma}}_0^{-1}) \bm{v} + \log |\tilde{\bm{\Sigma}}_1^{-1} \tilde{\bm{\Sigma}}_0|
\end{align}
where
\begin{align}
    \bm{v}&\triangleq \bm{y}^{\textnormal{comm}}_W -\alpha_W \hat{h}_W \bm{x}_A\\
    \tilde{\bm{\Sigma}}_0&\triangleq \mathbb{E}[\bm{v} \bm{v}^H|H_0,\widetilde{H}_0]=\sigma_W^2 \bm{I}\\
    \tilde{\bm{\Sigma}}_1&\triangleq \mathbb{E}[\bm{v} \bm{v}^H|H_0,\widetilde{H}_1]=(\alpha_W^2 |\hat{h}_W|^2 \Lambda_T+\sigma_W^2) \bm{I}
\end{align}

Thus, $\ell(\bm{y}^{\textnormal{comm}}_W|H_0)$ can be rewritten as
\begin{align}
    \ell(\bm{y}^{\textnormal{comm}}_W|H_0) &= \frac{1}{2}\left(\frac{1}{\sigma_W^2}-\frac{1}{\alpha_W^2 |\hat{h}_W|^2 \Lambda_T+\sigma_W^2}\right) \|\bm{v}\|_2^2 \notag\\
    & \hspace{1em}+ n \log\left(\frac{\sigma_W^2}{\alpha_W^2 |\hat{h}_W|^2 \Lambda_T+\sigma_W^2}\right) \label{eq: llr}
\end{align}
Plugging Eq.~\eqref{eq: llr} into Eq.~\eqref{eq: llr test} shows that Willie's optimal test is of the form
\begin{align}
    \frac{1}{n}\|\bm{y}^{\textnormal{comm}}_W-\alpha_W \hat{h}_W\bm{x}_A\|_2^2 \underset{\widetilde{H}_0}{\overset{\widetilde{H}_1}{\gtrless}}\tau
\end{align}
for some $\tau>0$.

Now, we focus on Willie's choice of the test threshold $\tau$. Let $\tau^\dagger$ denote Willie's optimal detection threshold. Formally, let
\begin{align}
    \tau^\dagger &\triangleq \argmin\limits_{\tau>0} \mathbb{P}^\dagger_F+\mathbb{P}^\dagger_M
\end{align}
where
\begin{align}
    \mathbb{P}^\dagger_F &\triangleq \Pr(T(\bm{y}^{\textnormal{comm}}_W)>\tau|\widetilde{H}_0,H_0)\\
    \mathbb{P}^\dagger_M &\triangleq \Pr(T(\bm{y}^{\textnormal{comm}}_W)<\tau|\widetilde{H}_1,H_0)
\end{align}
denote Willie's probability of false alarm and that of missed detection under $H_0$, respectively.

Although omitted here for brevity, one could verify the convexity of $\mathbb{P}^\dagger_F+\mathbb{P}^\dagger_M$ in $\tau$. Thus we need to find $\tau^\dagger$ such that
\begin{align}
    \frac{\partial (\mathbb{P}^\dagger_F+\mathbb{P}^\dagger_M)}{\partial \tau}\Big|_{\tau=\tau^\dagger} = 0
\end{align}

Note that $\mathbb{P}^\dagger_F$ and $\mathbb{P}^\dagger_M$ could be written as
\begin{align}
    \mathbb{P}^\dagger_F &= \Pr\left(\Lambda_Z>\tau\right)\\
    \mathbb{P}^\dagger_M &= \Pr\left(\alpha_W^2 |\hat{h}_W|^2 \Lambda_T+\Lambda_Z<\tau\right)
\end{align}
where $\Lambda_Z \triangleq \frac{1}{n}\|\bm{z}^{\textnormal{comm}}_W\|_2^2$. Hence, we obtain
\begin{align}
    \mathbb{P}^\dagger_F + \mathbb{P}^\dagger_M &= 1-\Pr\left(\tau-\alpha_W^2 |\hat{h}_W|^2 \Lambda_T\le \Lambda_Z\le \tau\right)
\end{align}

Note that $V\triangleq \frac{2n}{\sigma_W^2}\Lambda_Z$ follows a $\chi^2(2n)$-distribution. Hence, we have
\begin{align}
    \mathbb{P}^\dagger_F + \mathbb{P}^\dagger_M &= 1-\Pr\left(\frac{2n(\tau-\alpha_W^2 |\hat{h}_W|^2 \Lambda_T)}{\sigma_W^2}\le V \le \frac{2 n \tau}{\sigma_W^2}\right)\\
    &= 1- \int_{\frac{2n(\tau-\alpha_W^2 |\hat{h}_W|^2 \Lambda_T)}{\sigma_W^2}}^{\frac{2 n \tau}{\sigma_W^2}} \frac{v^{n-1} e^n}{2^n \Gamma(n)} dv \label{eq: P_F + P_M with no scaling}
\end{align}
The Leibniz integral rule~\cite{flanders1973differentiation} yields
\begin{align}
    &\frac{\partial (\mathbb{P}^\dagger_F+\mathbb{P}^\dagger_M)}{\partial \tau}\notag\\ &= \frac{n^n \Big((\tau-\alpha_W^2 |\hat{h}_W|^2 \Lambda_T) e^{\frac{-n(\tau-\alpha_W^2 |\hat{h}_W|^2 \Lambda_T)}{(n-1)\sigma_W^2}}-\tau e^{-\frac{n \tau}{(n-1)\sigma_W^2}}\Big)^{n-1}}{\Gamma(n)\sigma_W^n}
\end{align}
Solving 
\begin{align}
    (\tau^\dagger-\alpha_W^2 |\hat{h}_W|^2 \Lambda_T) e^{\frac{-n(\tau^\dagger-\alpha_W^2 |\hat{h}_W|^2 \Lambda_T)}{(n-1)\sigma_W^2}}-\tau^\dagger e^{-\frac{n \tau^\dagger}{(n-1)\sigma_W^2}}=0
\end{align}
we find Willie's optimal threshold $\tau^\dagger$ to be
\begin{align}
    \tau^\dagger = \alpha_W^2|\hat{h}_W|^2 \Lambda_T \frac{e^{\frac{n}{n-1}\frac{\alpha_W^2|\hat{h}_W|^2\Lambda_T}{\sigma_W^2}}}{e^{\frac{n}{n-1}\frac{\alpha_W^2|\hat{h}_W|^2\Lambda_T}{\sigma_W^2}}-1}.
\end{align}
\qed

\subsection{Proof of Lemma~\ref{lem: covert transmission}}\label{proof: covert transmission}
First, we rewrite $\mathbb{P}^{(2)}_F$ and $\mathbb{P}^{(2)}_M$ as
    \begin{align}
    \mathbb{P}^{(2)}_F &= \Pr\left(\epsilon^2 \alpha_W^2 |h_W|^2 \Lambda_A+\Lambda_Z>\tau(\epsilon)\right)\\
    \mathbb{P}^{(2)}_M &=\Pr\left(\epsilon^2 \alpha_W^2 |h_W|^2 \Lambda_A+\alpha_W^2 |h_W|^2 \Lambda_T+\Lambda_Z<\tau(\epsilon)\right)
\end{align}

    Next, assume Eq.~\eqref{eq: covertness threshold 1} holds, let
    \begin{align}
        \Delta_1 &\triangleq \frac{\epsilon^2 \alpha_W^2 |h_W|^2 \Lambda_A + \sigma_W^2 - \tau(\epsilon)}{\sigma_W^2}.
    \end{align}

Then, we obtain
\begin{align}
    1 - (\mathbb{P}^{(2)}_F+\mathbb{P}^{(2)}_M)&\le 1-\mathbb{P}^{(2)}_F\\
    &= \Pr\left(\epsilon^2 \alpha_W^2 |h_W|^2 \Lambda_A+\Lambda_Z\le \tau(\epsilon)\right)\\
    &= \Pr\left(\Lambda_Z\le \tau(\epsilon)-\epsilon^2 \alpha_W^2 |h_W|^2 \Lambda_A\right)
\end{align}
Once again, letting $V\triangleq \frac{2n}{\sigma_W^2}\Lambda_Z$ and noticing $V\sim\chi^2(2n)$, we get
\begin{align}
    1 - (\mathbb{P}^{(2)}_F+\mathbb{P}^{(2)}_M)&\le \Pr\left(V\le \frac{2n}{\sigma_W^2}(\tau(\epsilon)-\epsilon^2 \alpha_W^2 |h_W|^2 \Lambda_A)\right)\\
    &= \Pr\left(V\le 2n(1-\Delta_1)\right)\\
    &\le e^{-\frac{1}{2}n \Delta_1^2} \label{eq: chi-squared tail bound 1}
\end{align}
where Eq.~\eqref{eq: chi-squared tail bound 1} follows from \cite[Lemma 1]{laurent2000adaptive}. Thus, Eq.~\eqref{eq: covertness threshold 1} implies that $\mathbb{P}^{(2)}_F+\mathbb{P}^{(2)}_M\to 1$ exponentially as $n\to\infty$.

Finally, we assume Eq.~\eqref{eq: covertness threshold 2} holds and let
\begin{align}
    \Delta_2 &\triangleq \frac{\tau(\epsilon)-\epsilon^2 \alpha_W^2 |h_W|^2 \Lambda_A-\alpha_W^2 |h_W|^2 \Lambda_T-\sigma_W^2}{\sigma_W^2}>0
\end{align}

Then, 
\begin{align}
    1-(\mathbb{P}^{(2)}_F+\mathbb{P}^{(2)}_M) &\le 1- \mathbb{P}^{(2)}_M\\
    &= \Pr\left(\Lambda_Z\ge \sigma_W^2(1+\Delta_2)\right)\\
    &= \Pr(V\ge 2n(1+\Delta_2))\\
    &\le e^{-n(1+\Delta_2+\sqrt{1+2 \Delta_2})}\label{eq: chi-squared tail bound 2}
\end{align}
where Eq.~\eqref{eq: chi-squared tail bound 2} follows from \cite[Lemma 1]{laurent2000adaptive}. Thus, Eq.~\eqref{eq: covertness threshold 2} implies that $\mathbb{P}^{(2)}_F+\mathbb{P}^{(2)}_M\to 1$ as $n\to\infty$. 

The proof of the converse part follows the same steps, and hence is omitted for brevity.
\qed

\subsection{Proof of Lemma~\ref{lem: square root law}}\label{proof: square root law}
We first argue that when $\epsilon=0$, Tom cannot communicate at a non-vanishing power $\Lambda_T$, by showing that when $\epsilon=0$
    \begin{align}
        \tau(0) \in(
         \sigma_W^2,  \alpha_W^2 |h_W|^2 \Lambda_T + \sigma_W^2)
    \end{align}
for any $\Lambda_T>0$ whenever $\alpha_W^2$ and $|h_W|^2$ are positive. Hence, we argue that when $\epsilon=0$, Eq.~\eqref{eq: covertness threshold 1} or Eq.~\eqref{eq: covertness threshold 2} cannot be satisfied.

Note that when $\epsilon=0$, we have
    \begin{align}
        \tau(0) &=\alpha_W^2|h_W|^2 \Lambda_T \frac{e^{\frac{\alpha_W^2|h_W|^2\Lambda_T}{\sigma_W^2}}}{e^{\frac{\alpha_W^2|h_W|^2\Lambda_T}{\sigma_W^2}}-1}
    \end{align}
    Let 
    \begin{align}
        \beta_T&\triangleq \alpha_W^2|h_W|^2 \Lambda_T\\
        f(\beta_T)&\triangleq \beta_T \frac{e^{\beta_T/\sigma_W^2}}{e^{\beta_T/\sigma_W^2}-1}.
    \end{align}
    It is straightforward to show that 
    \begin{align}
        \lim\limits_{\Lambda_T\to 0} f(\beta_T) &= \sigma_W^2\\
        f^\prime(\beta_T)>0 \hspace{0.5em}&\hspace{0.5em} \forall \Lambda_T>0.
    \end{align}
    Thus, $f(\beta_T)>\sigma_W^2$, $\forall \Lambda_T>0$. In turn, we obtain ${\tau(0)>\sigma_W^2}$, $\forall \Lambda_T>0$.

    Similarly, let 
    \begin{align}
        g(\beta_T)\triangleq f(\beta_T)-\beta_T.
    \end{align}
    Again, one can check that 
    \begin{align}
        \lim\limits_{\Lambda_T\to0} g(\beta_T) &= \sigma_W^2\\
        g^\prime(\beta_T)<0 \hspace{0.5em} & \hspace{0.5em} \forall \Lambda_T>0.
    \end{align}
    Thus, $\tau(0)<\sigma_W^2+\alpha_W^2 |h_W|^2 \Lambda_T$, $\forall \Lambda_T>0$. Thus we have showed that Tom cannot covertly communicate at a non-vanishing power.

Now we suppose that Tom communicates at a vanishing power $\Lambda_T=o_n(1)$. More specifically, to prove the achievability result, suppose that $\Lambda_T = \mathcal{O}(n^{-\nicefrac{1}{2}})$. 

Considering the first-order MacLaurin series expansion of $\tau(0)$ with respect to $\Lambda_T$, it is straightforward to obtain
    \begin{align}
        \tau(0) & = \sigma_W^2 + \frac{\beta_T}{2}+ \frac{\beta_T^2}{12 \sigma_W^2}- \mathcal{O}(\beta_T^4)
    \end{align}
    Hence, for sufficiently large $n$, we have
    \begin{align}
        \tau(0) &> \sigma_W^2 +\frac{\beta_T}{2} \label{eq: srl eq1}\\
        \tau(0) &< \sigma_W^2 +\frac{3 \beta_T}{4} \label{eq: srl eq2}
    \end{align}
    Recalling Eq.~\eqref{eq: P_F + P_M with no scaling}, we have
    \begin{align}
        1-\mathbb{P}^\dagger_F - \mathbb{P}^\dagger_M 
    &= \int_{\frac{2n(\tau(0)-\beta_T)}{\sigma_W^2}}^{\frac{2 n \tau(0)}{\sigma_W^2}} \frac{v^{n-1} e^{-v/2}}{2^n \Gamma(n)} dv \label{eq: 1 - P_F - P_M no scaling}
    \end{align}
    Since the integrand is monotonically increasing for sufficiently large $n$, the integral can be bounded above by any right Riemann sum. Thus,
    \begin{align}
        1-\mathbb{P}^\dagger_F - \mathbb{P}^\dagger_M 
    &\le \frac{n^{n+1}}{n!}\frac{\beta_T}{\sigma_W^{2n}} (\tau(0))^{n-1}\exp\left(\frac{2n\tau(0)}{\sigma_W^2}\right)
    \end{align}
    Applying Stirling's approximation~\cite[Chapter 3.2]{cormen2022introduction}, we get
    \begin{align}
        1-\mathbb{P}^\dagger_F - \mathbb{P}^\dagger_M 
    &\le \frac{\sqrt{n}\beta_T}{\sqrt{2 \pi} \tau(0)} \left(\frac{\tau(0)}{\sigma_W^2}\right)^n \exp\left(-n\left(\frac{\tau(0)}{\sigma_W^2}-1\right)\right)\label{eq: srl eq3}\\
    &\le \frac{\sqrt{n}\beta_T}{\sqrt{2 \pi} \sigma_W^2} \left(1+\frac{\beta_T}{2\sigma_W^2}\right)^n \exp\left(-n\frac{\beta_T}{2\sigma_W^2}\right) \label{eq: srl eq4}
    \end{align}
    where Eq.~\eqref{eq: srl eq4} follows from Eq.~\eqref{eq: srl eq1} and the fact that the RHS of Eq.~\eqref{eq: srl eq3} is decreasing in $\beta_T$.
    
    Note that if $\Lambda_T=o(n^{-\nicefrac{1}{2}})$ and in turn $\beta_T=o(n^{-\nicefrac{1}{2}})$, by taking the MacLaurin series of the RHS of Eq.~\eqref{eq: srl eq4}, we get
    \begin{align}
        1-\mathbb{P}^\dagger_F - \mathbb{P}^\dagger_M 
    &\le \frac{7}{8\sqrt{2 \pi}}\beta_T \sqrt{n}
    \end{align}
    Hence, $\lim\limits_{n\to\infty} \mathbb{P}^\dagger_F + \mathbb{P}^\dagger_M =1$ if $\Lambda_T=o(n^{-\nicefrac{1}{2}})$. 

    Now, suppose that $\sqrt{n}\Lambda_T = C$ for some constant $C\in\mathbb{R}$. Then, by taking the limit of Eq.~\eqref{eq: srl eq4}, we obtain
    \begin{align}
        \lim\limits_{n\to\infty} 1-\mathbb{P}^\dagger_F - \mathbb{P}^\dagger_M &\le \frac{\alpha_W^2 |h_W|^2 C}{\sqrt{2\pi}\sigma_W^2} \exp\left(-\frac{\alpha_W^4 |h_W|^4 C^2}{8 \sigma_W^4}\right)
    \end{align}
    Thus, for any $\delta_2>0$, Tom could select $C$ small enough to satisfy the covertness constraint $\lim\limits_{n\to\infty} 1-\mathbb{P}^\dagger_F - \mathbb{P}^\dagger_M \le \delta_2$.

    Next, we prove the converse result. Let $\Lambda_T=\omega(n^{-\nicefrac{1}{2}})$. Recall that
    \begin{align}
        \mathbb{P}^\dagger_F &= \Pr\left(\Lambda_Z>\tau(0)\right)\\
        \mathbb{P}^\dagger_M &= \Pr\left(\Lambda_Z+\alpha_W^2 |h_W|^2 \Lambda_T <\tau(0) \right)
    \end{align}
    where $\Lambda_Z\triangleq \frac{1}{n}\|\bm{z}_W^{\textnormal{comm}}\|_2^2$. Once again, let $V\triangleq \frac{2n}{\sigma_W^2}\Lambda_Z$. Using Eq.~\eqref{eq: srl eq1}, we obtain 
    \begin{align}
        \mathbb{P}^\dagger_F &\le \Pr\left(V> 2n\left(1+\frac{\alpha_W^2 |h_W|^2 \Lambda_T}{2\sigma_W^2}\right)\right)\\
        &\le \exp\left(-\frac{\alpha_W^4 |h_W|^4}{24} n \Lambda_T^2\right)\label{eq: srl eq5}
    \end{align}
    where Eq.~\eqref{eq: srl eq5} follows from~\cite[Lemma 1]{laurent2000adaptive}. Hence $\Lambda_T=\omega(n^{-\nicefrac{1}{2}})$ implies that $\lim\limits_{n\to\infty} \mathbb{P}^\dagger_F = 0$.
    
    Now, we focus on $\mathbb{P}^\dagger_M$. Note that Eq.~\eqref{eq: srl eq2} yields
    \begin{align}
        \mathbb{P}^\dagger_M &= \Pr\left(\frac{\sigma_W^2}{2n} V < \tau(0)-\alpha_W^2 |h_W|^2 \Lambda_T\right)\\
        &<\Pr\left(\frac{\sigma_W^2}{2n} V < \sigma_W^2-\frac{\alpha_W^2 |h_W|^2 \Lambda_T}{4}\right)\\
        &\le \exp\left(-\frac{\alpha_W^4 |h_W|^4}{16 \sigma_W^4} n \Lambda_T^2\right)\label{eq: srl eq6}
    \end{align}
    where Eq.~\eqref{eq: srl eq6} follows from \cite[Lemma 1]{laurent2000adaptive}. Thus, $\Lambda_T=\omega(n^{-\nicefrac{1}{2}})$ implies $\lim\limits_{n\to\infty}\mathbb{P}^\dagger_M=0$. This concludes the proof.
    \qed
}
\end{document}